\documentstyle[aps,epsfig]{revtex}

\begin{document}

\draft

\title{Vortices in Ginzburg-Landau billiards}
\author{E.~Akkermans and K.~Mallick\cite{kirone}\\ Department of Physics,
 Technion, 32000 Haifa, Israel}

\maketitle

\begin{abstract}
We present an analysis of the Ginzburg-Landau equations for the description of 
a two-dimensional superconductor in a bounded domain. Using the properties of 
a special integrability point of these equations which allows vortex solutions,
 we obtain a closed expression for the energy of the superconductor.
 The role of the 
boundary of the system is to provide a selection mechanism for the
 number of vortices.
 A geometrical interpretation of these results is presented and they 
are applied to the analysis of the magnetization recently measured 
on small superconducting disks. Problems related to the interaction
 and nucleation 
of vortices are discussed.
\end{abstract}
 

\def\real{{\rm I\kern-.2em R}}
\def\complex{\kern.1em{\raise.47ex\hbox{
	    $\scriptscriptstyle |$}}\kern-.40em{\rm C}}
\def\integer{{\rm Z\kern-.32em Z}}
\def\pinteger{{\rm I\kern-.15em N}}

\newpage

\section {Introduction}

This article is devoted to the study of the existence and stability of vortex 
solutions of the Ginzburg-Landau equations in finite  two-dimensional domains
 with boundaries. 
 For   the case of the infinite plane,
 a  large amount of work has been devoted to such questions,
 motivated by the ubiquitous character of the Ginzburg-Landau
 equations either in the study of superconductors or for the
 Abelian Higgs model.
 It is known \cite{bogo,sarma,taubes} that for the infinite system
 the Ginzburg-Landau energy 
functional has a lower bound which is saturated  for a special choice
 of the parameters (which corresponds to the limiting case
   between type I and type II superconductors).
 The minimizing  equations at this special point, also called
 the  dual  point,
 are first order 
differential equations and they admit  solutions with  vortices.
  These different vortex solutions are classified  by an integer of 
topological origin, namely the winding number of the order parameter which,
roughly speaking, counts the number of vortices.

  Although the  
Ginzburg-Landau equations in finite domains
 have  been investigated \cite{bethuel}, there is
 no  generalization so far of the results concerning the dual 
  point for finite two-dimensional domains (that we shall also  call
  hereafter `billiards').
In this paper, we  study such a generalization;  besides  a 
 theoretical interest, our motivation has been triggered
 by a set of new experimental results 
obtained on small  aluminium disks \cite{geim} in a
  mesoscopic  regime where the 
radius $R$ is comparable with both the coherence length  $\xi$
 and the London penetration
 length $\lambda.$ There, the magnetization, 
as a function of the
 applied magnetic field,  presents  a series of jumps
 with an overall shape  reminiscent of type-II superconductors.
 Such a   behaviour is very different from that  of
  macroscopic systems: in the large, aluminium is 
 a genuine type-I superconductor.
 These experiments  were  first analyzed in 
the framework of 
the linearized Ginzburg-Landau equations \cite{zwerger}. 
 Although this
 approach explains
   some of the  observed features, it   fails to provide a satisfactory
 quantitative
 agreement, and can not account for the
 large  spatial variations of the magnetic
  field inside the sample (e.g. vortices).
  Numerical solutions \cite{deo} of the Ginzburg-Landau equations 
  give a much better description of this phenomenon, and 
  emphasize  the importance of
 the nonlinear term. However,
 in this context,  no analytical results  are  available for the
  full Ginzburg-Landau equations. Using our results on the dual  point,
 we shall derive an analytical 
 expression for the free energy and  for the magnetization
 of a mesoscopic superconductor as a function of the applied
 magnetic field.

The plan of this paper is as follows. In the section 2, we shall describe the 
main features 
and the known results about the dual point and the
lower energy bound for an infinite system. 
Then, in section 3, we shall study the case of a finite
 domain and derive  an expression 
of the free energy which, in addition
 to the previous  lower bound, includes a 
contribution of the boundary. Section 4 contains
 a geometrical interpretation  of these results. In section 5, 
 our results   are applied  to  analyze   
 experimental data  on small superconducting disks.
 In the conclusion, we propose some extensions of our work and
 a scenario for the nucleation of vortices at the boundary.

\section {The dual point of Ginzburg-Landau equations in an infinite system}

   From now on, we study a superconducting billiard
 within  the framework of the Ginzburg-Landau equations 
 (this assumes that both the order 
parameter and the vector potential have a slow spatial variation).
 The expression of the Ginzburg-Landau
 energy density $a$ is given by
\begin{equation}
a= {a_0} + {a_2}|{\psi}{|^2} + {a_4}|{\psi}{|^4} +{a_1} 
 |({\vec{\nabla}}- i{{2e} \over {\hbar c}}
{\vec A}) \psi{|^2} + {{B^2} \over {8\pi}}
\end{equation}
where $\psi = |\psi| {e^{i \chi}}$ is the complex-valued  order parameter,
 $B$ is  the magnetic field  
 and the $a_i$'s are real 
parameters. The coherence length and the London penetration
 length  are related to these parameters
 as follows  \cite{degennes}: 
 $\xi^2 =  {{a_1} \over {|a_2|}}$ and 
  ${\lambda^2} = 
{{\sqrt 2} \over {4\pi}} ({\hbar c \over {2e}}{)^2} {{a_4}
 \over {{a_1}|a_2|}} $,  so that we obtain for
 the   dimensionless  free energy $\cal F$  
\begin{equation}
{\cal F} = {\int_{\Omega}} {1 \over 2} |B{|^2} + {\kappa^2}
 |1 - |\psi{|^2}{|^2} + 
|({\vec{\nabla}} - i{\vec A})\psi {|^2}
\end{equation}
  where $\psi$  is measured in units of ${\psi_0} =
 \sqrt {{|{a_2}|} \over {2{a_4}}}$, 
 $B$ in units of  
${ {\phi_0} \over {4 \pi \lambda^2}}$, where ${\phi_0} = { hc \over 2e}$
is the quantum of flux.  
 The lengths are measured  in units of 
$\lambda { \sqrt 2} $ (the numerical factor $\sqrt 2$ is for
 further convenience).
The ratio  $\kappa = {\lambda \over \xi}$ is the only free parameter
 in (2) and it determines,
 in the limit of an  
infinite system, whether the sample is a type-I or type-II superconductor
 \cite{degennes}. The integral is 
over the two-dimensional domain $\Omega$ of the  superconducting sample.
 
The Ginzburg-Landau equations for the order parameter $\psi$ and for the
 magnetic 
field ${\vec B}= {\vec \nabla} \times {\vec A} $  are obtained
 from a variation of $\cal F$:
\begin{eqnarray}
   ({\vec{\nabla}} - i{\vec A})^2 \psi & = &  2{\kappa}^2 \psi
(1 - |\psi{|^2})   \nonumber \\
    {\vec \nabla} \times {\vec B} & = & 2 {\vec \jmath}
\end{eqnarray}
 where the current density ${\vec \jmath}$ is given by 
\begin{equation}
{\vec \jmath}= Im( {{\psi^*}} {\vec \nabla} \psi ) - |\psi{|^2}{\vec A}
\label{currentdens}
\end{equation}

  The Ginzburg-Landau equations  are 
nonlinear second order differential equations 
 and  their  solutions are not known 
 except 
 for some particular cases.
 However,  for the special value $\kappa = 
{1 \over \sqrt 2}$,  the equations 
for $\psi$ and ${\vec A}$  can be reduced to  first 
order differential equations. This special point was first
used   by Sarma \cite{sarma} in his  discussion 
of type-I vs type-II superconductors and then identified by
 Bogomol'nyi \cite{bogo} in 
the more general context of stability and integrability of classical
 solutions of some quantum field 
theories. 
  We now   review  some of these  properties of
 the Ginzburg-Landau free energy at  the dual  point. We shall 
 use  the following identity true 
for two dimensional systems
\begin{equation} 
|({\vec{\nabla}} - i{\vec A})\psi {|^2}= |{\cal D} \psi {|^2} +
 {\vec \nabla} \times {\vec \jmath} + B |\psi{|^2}
\label{ident}
\end{equation}
where ${\vec \jmath}$
is the current density and the operator $\cal D$ is 
defined as  ${\cal D} = \partial_x + i\partial_y -i(A_x + iA_y)$. 
  At the dual  point,  
$\kappa = {1 \over \sqrt 2},$ the
 expression (2) for   $\cal F$  can be rewritten using the identity
 (\ref{ident}) as follows 
\begin{equation}
{\cal F} = {\int_{\Omega}} {1 \over 2} | B - 1 + |\psi{|^2}{|^2} +
  |{\cal D} \psi {|^2}
 + {\oint_{\partial \Omega}} ({\vec\jmath} + {\vec A}).{\vec dl}
 \label{identitebog}
\end{equation}
where the last integral over the boundary ${\partial \Omega}$ of the
 system results from  Stokes theorem.

 For an infinite system  we impose, as in  \cite{bogo},
 that the system is superconducting at large distance, 
  i.e. $|\psi| \to 1$ and
 ${\vec \jmath} \to 0$  at infinity.
  The boundary term in (\ref{identitebog}) then becomes
  \begin{equation}
  {\oint_{\partial \Omega}} ({\vec \jmath} + {\vec A}).{\vec dl} =
  {\oint_{\partial \Omega}} 
  ( { {\vec\jmath} \over  {|\psi|^2}} 
  + {\vec A}).{\vec dl} \
 \label{fluxoid}
 \end{equation}
  This last integral is known as  the  London
fluxoid. It is    quantized and 
 using (\ref{currentdens}) one shows that it is equal to
 $\oint_{\partial\Omega} {\vec \nabla}\chi.{\vec dl}= 2 \pi n $,
 where $\chi$ is the phase of the order parameter.
  The integer $n$ is 
   the winding number of the order parameter $\psi$ and as such is a
 topological characteristic of the system. 
   The     extremal values of ${\cal F}$, 
 namely ${\cal F} = 2 \pi n$,   are  obtained
 when the bulk  integral in (\ref{identitebog}) vanishes identically, i.e. 
 when the two
    Bogomol'nyi \cite{bogo}
  equations  are  satisfied:
\begin{eqnarray}
 {\cal D} \psi & = &  0 
   \nonumber  \\
B & = &  1 - |\psi{|^2}
\label{equaBogo}
\end{eqnarray}   
 These two 
equations can be decoupled   and  one obtains that
   $|\psi|$ is  a  solution
 of the second order
  nonlinear equation 
\begin{equation}
{\nabla^2} ln |\psi{|^2} = 2 ( |\psi{|^2} -1)
\label{Liouville}
\end{equation}
 This equation  is  related  to  the Liouville equation.
 The  set of equations (\ref{equaBogo},\ref{Liouville})
  has been  obtained
 without
 any assumption on 
the nature of the magnetic field and  appears
 in various other situations,  
  e.g. Higgs \cite{taubes}, Yang-Mills
 \cite{witten} and
 Chern-Simons \cite{jackiw} field theories.  It 
was proven that these equations  admit families of vortex solutions 
\cite{taubes}. 
 For infinite systems,
 it can be shown  that each vortex carries
 one flux quantum and that the winding number  
 $n$ is equal to  the number of vortices in the
 system. However,  for an infinite system
 there is no mechanism to  select the value of $n$, which
 only plays the role of a classifying parameter. 
It will be  precisely the role of the 
boundary of a finite system to introduce such a  selection mechanism
 and to determine $n$, according to the applied
 magnetic field.
   
\section {The  finite size system}

 From  now on, we  shall  study finite size systems
 in an external magnetic field.
The question then arises to know if 
such systems  can sustain stable vortex solutions and 
how they  behave  as a function of the applied field.
A simplified version,
 without applied magnetic field,
 was extensively studied by 
Bethuel et al. \cite{bethuel}.
 In this work  the  mechanism for vortex creation is 
based on  Dirichlet boundary conditions of the type $\psi = f$ on 
$\partial \Omega$ and where $f$ is a complex function of degree $n$.
 In the London limit, namely 
$\kappa \to \infty$, $|\psi|$ is 1 almost everywhere
 but, because of the degree $n$ imposed
 on the boundary,
 $\psi$  must vanish $n$ times in the bulk therefore leading to vortices. 
 Moreover, numerical simulations
 of the Ginzburg-Landau equations 
  for a long and thin
 parallepiped in a uniform magnetic field \cite{argentins}
 show the existence of stationary vortex solutions
 whose number depends on the applied  magnetic field.
 These simulations  then indicate that the physical picture  derived for
$\kappa = {1 \over \sqrt 2}$ 
remains valid  for quite a large range of values of  $\kappa$,
 and  that the corresponding  change of free  energy is  small 
(see also \cite{rebbi}). Indeed,
 for finite systems  their size $R$  
 becomes relevant as a new length scale so that $\kappa$
 is no more the only dimensionless
  parameter, and  its value does not alone  control the physics
 at  the mesoscopic scale   as it does  for
 infinite systems.
 We shall  study the case  $\kappa = {1 \over \sqrt 2}$, i.e. the dual
 point,  and extend the previous approach to a system with finite size where
  boundary effects are important.

 In  a finite system, there are in general
 non-zero  edge currents
 and the order parameter is not equal to
 1 on the boundary. Hence, the identification of the boundary
 integral in (\ref{identitebog}) with the fluxoid (\ref{fluxoid})
 is not possible anymore, and the 
free energy can not be minimized
 just by imposing Bogomol'nyi equations (\ref{equaBogo}).
 However, the  currents
 on the boundary of the  system screen the external magnetic field and 
therefore produce a magnetic moment (a circulation)  
opposite to the direction of the field, 
 whereas  vortices in the bulk of the system produce a
 magnetic
 moment along the direction of 
the applied field.
 Hence  currents in the bulk  circulate  in a direction opposite
 to those  at the boundary.
 If one assumes cylindrical symmetry, the current density
 $\vec\jmath$   has only 
 an azimuthal component, with  opposite  signs  in the bulk
 and on the edge of  the system 
 (the  radial component of  ${\vec\jmath}$ is zero     
 since  ${\vec\jmath}$ is  divergenceless).
 Thus,  there exists  a circle $\Gamma$  on which  $\vec\jmath$ vanishes
 \cite{remark}.
 This allows us to separate  the domain $\Omega$ into two concentric subdomains
 $\Omega = {\Omega_1}
\cup {\Omega_2}$ 
 such that the boundary  $\partial \Omega_1$  is  the curve
 $\Gamma$ (see Fig.1). On $\partial \Omega_1$, the current density
  ${\vec \jmath}$ is zero, therefore one has 
\begin{equation}
{\oint_{\partial \Omega_1}} {\vec \jmath}.{\vec dl}  =
  {\oint_{\partial \Omega_1}} 
   { {\vec \jmath} \over  {|\psi|^2}}.{\vec dl} = 0 
 \label{petitbogo}
\end{equation}

 Thus  one deduces as above
  that Bogomol'nyi and Liouville equations are valid in the finite domain
$\Omega_1$ as in the case of the infinite plane.
 The existence of vortices in a finite domain such as
   $\Omega_1$  was  checked using  a  numerical solution
  \cite{devega} of (\ref{Liouville}):
  assuming cylindrical 
symmetry, one shows  that  $|\psi|$  vanishes as a power law 
  at  the center of the disk,
 hence  there is a vortex in  the center (more precisely a multi-vortex
 whose multiplicity is determined by the  exponent of the power law);
   moreover,  $|\psi|$ saturates 
very rapidly to a constant value close to one for lengths
 larger than $\lambda$.
 The same  conclusion can be reached  by 
 defining  $ f(r) = - ln |\psi{|^2}$ and linearizing 
 (\ref{Liouville})  around $|\psi| = 1$. 
Then, $f$ satisfies  ${\nabla^2} f = 2f$ whose general
 solution is 
$f(r) =  a {I_0}(r {\sqrt 2}) + b {K_0}(r {\sqrt 2})$.  From the behaviour of
 the Bessel functions $I_0$ and $K_0$, one obtains  that
for small $r$, $|\psi|$ vanishes as a power law
 and saturates rapidly to a constant of order one in a finite range of $r$'s
 (in units of $\lambda\sqrt2$).  

  The  magnetic flux $\Phi ({\Omega_1})$ 
 through $\Omega_1$ 
   is calculated, in units of the flux quantum  $\phi_0$,
 using the fluxoid  and (\ref{petitbogo}) so that  
 $$\Phi ({\Omega_1}) =   n - \oint_{\partial{\Omega_1}} {{{\vec \jmath}.{\vec 
dl}}
 \over {|\psi{|^2}}} =   n $$
 As before  $n$ 
  is  the 
winding number, i.e. 
 $\oint_{\partial{\Omega_1}} {\vec \nabla }\chi.{\vec dl} =
 2\pi n$, as well as  the number of vortices.   
 The free energy in $\Omega_1$  is 
\begin{equation}
{\cal F}({\Omega_1})  = 2 \pi n .
\label{free1}
\end{equation}

We consider now the contribution of  $\Omega_2$ to the free energy.
It is given by  (2) and can be rewritten using
the phase and the modulus of the order parameter $\psi$,  as
 \begin{equation}
{\cal F}(\Omega_2)={\int_{\Omega_2}}  (\nabla |\psi|{)^2}
 + |\psi{|^2} |{\vec{\nabla}}\chi - {\vec A}{|^2}
 + {{B^2} \over 2} + { {(1 - |\psi{|^2}{)^2}} \over 2}
\label{tran1}
\end{equation}
  We know, from  the London equation, that both 
the magnetic field and the vector potential
  decrease rapidly (exponentially) away from
 the boundary $\partial\Omega$ of the system over a
 distance of order 1 in  units of $\lambda {\sqrt2}$.
  Over the same distance, at the dual point,   $|\psi|$ saturates to unity. 
  One can  thus estimate the integral (\ref{tran1}) using an elementary
 version of the saddle-point method. 
 We assume  cylindrical symmetry, and we  neglect 
 the term $(\nabla |\psi|{)^2}$  on the boundary  because of 
 the   covariant Neumann boundary conditions at the 
 interface between a superconductor 
  and an insulator \cite{geim}. 
 We obtain for the free energy the expression  
\begin{equation}
 {\cal F}(\Omega_2) \simeq   \oint_{\partial \Omega}
  |\psi{|^2} |{\vec \nabla} \chi - {\vec {A}}{|^2} + 
 {{B}^2 \over 2} + { {(1 - |\psi{|^2}{)^2}} \over 2}
\label{tran2}
\end{equation}
 where the integral is now over the boundary of the system. 
To go further, we need to implement
 boundary conditions for  the magnetic field $ B(R)$ 
 and the vector potential $A(R)$.
 The  choice  $ B(R) =  {B_e}$, where $B_e$ is the external imposed field,
  corresponds  to the geometry of an infinitely long cylinder,
 where the flux lines are not distorted outside the system.
 This boundary condition is not 
 adapted to describe  a flat thin  disk.
 A more suitable choice  is provided by demanding
  $\phi = {\phi_e},$ which means that the total flux through the disk
 is identical to the flux of the external field, although flux lines are distorted
 by the superconducting sample. One can check, using numerical  simulations
 \cite{peetschw}, that this condition is well satisfied if
 the disk is thin enough.
 The boundary condition  $\phi = {\phi_e}$ implies that
  the vector potential
 is  identified by continuity
 to its external applied  value ${\vec {A_e}}$
 which  has only the  azimuthal component
  ${{\phi_e} \over {2 \pi R}}$.
 It should be noticed that
  the magnetic field ${\vec B}$ has a non monotonous  variation:
  it is low in the bulk, larger than $B_e$ near the edge of the system,
 because of the distortion of flux lines,
 and eventually equal to  its applied value 
 far  outside the system
  \cite{deopeeters}.

 Different choices of boundary conditions  will give rise
 to  different  limits  for a very large system (i.e. for $R \to \infty$). 
 The limit  of an infinitely long cylinder  
 corresponds to a superconducting bulk sample, whereas
 the limit of a thin disk, which is the case we consider here,
 corresponds to a superconducting thin sheet.

 The  formula  (\ref{tran2})  for  the free energy is similar to
  the Little-Parks expression
 \cite{degennes}
 for a quasi one-dimensional hollow cylinder in a uniform applied magnetic
 field.
 The minimization with respect to $|\psi|$ gives
 $1 - |\psi{|^2} = 
|{\vec \nabla} \chi - {\vec A}{|^2}$,
 such that  (\ref{tran2}) can be written as
\begin{equation}
{\cal F}(\Omega_2) =  \oint_{\partial\Omega} |{\vec \nabla}
 \chi - {\vec {A_e}{|^2} + 
{1 \over 2}{B^2} - {1 \over 2} |{\vec \nabla} \chi - {\vec {A_e}}{|^4}}
\label{tran3}
\end{equation}
Performing the integral over the boundary of the system, we obtain
\begin{equation}
{1 \over {2 \pi}}{\cal F}(\Omega_2) = {{ \lambda {\sqrt 2}} \over R}
 (n - {\phi_e} {)^2}
- {1 \over 2} ({{\lambda {\sqrt 2}} \over R}{)^3}(n - {\phi_e} {)^4}
\label{free2}
\end{equation}
 We have neglected the contribution of the  $B^2$ term, which is similar to
  the first term in the r.h.s. of (\ref{free2}) but smaller  by a factor
 of the order $(\lambda/R)^2$.
 The integer $n$ which appears in (\ref{free2})
 is the same as in (\ref{free1}), since  the 
order parameter $\psi$ is the same  function in both subdomains.
 The circulation of its phase $\chi$ (the winding 
number) counts the number of 
zeroes in the domain $\Omega$, i.e. the number of vortices.
 The thermodynamic Gibbs potential ${\cal G}$ of the system is obtained 
 from ${\cal F}(\Omega_1) + {\cal F}(\Omega_2)$ by a Legendre transformation 
 so that
\begin{equation}
{1 \over {2 \pi}} {\cal G}(n, {\phi_e}) = n + {{ \lambda {\sqrt 2}} \over R}
 (n - {\phi_e} {)^2}
- {1 \over 2} ({{\lambda {\sqrt 2}} \over R}{)^3}(n - {\phi_e} {)^4}
 - {{ 2 \lambda^2}
 \over {R^2}}{\phi_e}^2
\label{gibbs}
\end{equation}
This relation consists in a set of quartic
 functions indexed by the integer $n$. The minimum 
of the Gibbs potential is the envelop curve defined by the equation 
${{\partial {\cal G}} \over {\partial n}}{|_{\phi_e}} =0$, i.e.
 the system chooses its winding number 
$n$ in order to minimize $\cal G$. This provides a relation between 
the number $n$ of vortices in the system and the applied magnetic field 
$\phi_e$. 
In the limit of a large enough $ R \over \lambda$, the quartic
 term is negligible and the 
Gibbs potential reduces to a set of parabolas (Fig.2).
 The winding number $n$ is then given by the integer part 
\begin{equation}
n = [  { \phi_e}
 - {R \over {2 {\sqrt2}\lambda}} + {1 \over 2}  ] 
\end{equation}
 The magnetization 
   $ M= - {  {\partial {\cal G}}   \over  {\partial \phi_e}  },$ of the  system, is given by
\begin{equation}
- M = {{2{\sqrt 2} \lambda} \over R} ( {\phi_e} - n ) -
 {{ 4 \lambda^2} \over {R^2}} {\phi_e}
\label{magnet}
\end{equation}
For ${\phi_e}$ smaller that $R \over {2{{\sqrt 2}\lambda}}$,
 we have $n =0$  and 
$(- M)$  increases linearly with 
 the external flux. This corresponds 
to the London regime before the first vortex enters the system.
 The field $H_1$ at which the first vortex enters the system 
 corresponds to ${\cal G}(n=0)=
 {\cal G}(n=1)$, i.e. to 
\begin{equation}
  {H_1} = { {\phi_0} \over {2 \pi {\sqrt 2} R \lambda} }
 + { {\phi_0} \over {2 \pi  R^2} }
\end{equation}
 The subsequent 
vortices enter one by one for each crossing ${\cal G}(n+1)= {\cal G}(n)$; 
 this happens periodically in the applied field, with a period equal to
\begin{equation}
 {\Delta}H = {{\phi_0}
 \over { \pi  {R^2}}}
\end{equation}
 This gives rise to a discontinuity of the magnetization
 $ {\Delta}M = {{2 {\sqrt 2} \lambda} \over R }.$

There is a qualitative similarity between the results
 we derived using the Bogomol'nyi 
equations within the domain $\Omega_1$ and those obtained
 from a linearised version of the 
Ginzburg-Landau functional \cite{zwerger}. But the two approaches differ in 
their 
quantitative predictions due to the importance of the non linear
 term. An illustration of this 
is given in the next section.

\section {A geometrical interpretation}

 In an infinite system, one shows 
 using the boundary conditions 
$|\psi| \to 1$ and ${\vec \jmath} \to 0$  at infinity
 that equation  (\ref{fluxoid})
 implies the quantization of 
 the magnetic flux:
\begin{equation}
{\int_{R^2}} {\vec B}.d{\vec S} = n
\label{chern}
\end{equation}
  If one interprets $B$ as a curvature, this relation is analogous to the
 Gauss-Bonnet theorem, which states 
 that,  for  a compact manifold $M$  without boundary,
 the integral of the Gaussian  curvature $K$ over the surface
 is equal to 
 the Euler-Poincar\'e characterics $\chi$  of the manifold,
 which is a  topological invariant integer:
\begin{equation}
 \int_{M} {K} = \chi
\label{gaussbonnet}
\end{equation}
 This result is in fact  more than an analogy 
and at the dual point  the Ginzburg-Landau functional 
has a useful geometrical interpretation that we shall now highlight
 \cite{am1}. 

The Ginzburg-Landau functional (2) for the
 energy of a superconductor corresponds to a U(1) 
gauge symmetry. For that symmetry,  one can identify 
an abelian one-form connexion and
 a two-form curvature $\Omega$ respectively given by the
 vector potential $\vec A$ and by the magnetic field $\vec B$.
 The complex order parameter 
$\psi$ is a section of the U(1) fiber
 and the basis manifold is the infinite plane. 
 The above boundary conditions  allow to map
 the plan   onto  a compact manifold without boundary, namely   the sphere.
 Topological invariants of the problem are obtained from 
the Chern-Weil invariant polynomial
 $P(\Omega) = det (1 + {i \over {2 \pi}} \Omega)$ (see e.g. \cite{eguchi})
For the U(1) bundle over a sphere
there is only one Chern class, $ \Omega (= B)$. The integral 
of that Chern class
 over the basis manifold is a  topological invariant integer
called a  Chern number and 
 is precisely  given by (\ref{chern}). This Chern number plays
 a role similar to $\chi$.
 At  the dual  point,
 the energy is  ${\cal F}
 = \int  B  =  2 \pi n;$ this fact   can be translated 
in geometrical terms by  saying that the extremal  free energy
 is identical to a   topological invariant of the 
problem namely its Chern number.

When the basis manifold $M$ has a boundary $\partial M$
 which is not a geodesic,
 the integral of its curvature 
is neither an integer nor a topological invariant. 
  The Gauss-Bonnet theorem is then generalized  by
 adding a boundary term so that the 
Euler-Poincar\'e characteristics $\chi$ is now given by the relation
\begin{equation}
\chi = {1 \over {2 \pi}} {\int_M} K dS +
 {1 \over {2 \pi}}  \oint_{\partial M} {k_g} dl
\label{gb}
\end{equation}
where $K$ and $k_g$ are respectively 
the  Gaussian curvature   of the manifold 
 and the geodesic curvature
  of the boundary. 
 A  similar  result  holds for the U(1) problem in 
 a bounded domain (or in  an infinite domain with  
  boundary conditions different from those chosen above).
  In that case the relation equivalent to (\ref{gb}) is given 
by the fluxoid relation (\ref{fluxoid}):
\begin{equation}
 n = 
{1 \over {2 \pi}} {\int_\Omega} {\vec B}.d{\vec S} +
\oint_{\partial \Omega} {{\vec  \jmath} \over {|\psi{|^2}}}.d{\vec l}
\end{equation}
 As before,  $B$ is  the curvature of the connexion 
  and the current density 
${{\vec  \jmath} \over {|\psi{|^2}}}$ plays here
 the role of a  geodesic curvature \cite{am1}.
The expression  obtained in (\ref{free2})
 for the Bogomol'nyi free energy of  a system
 with a boundary can be rewritten as:
\begin{equation} 
 {\cal F} = \int_{\Omega}  B   + \int_{\partial\Omega}
\eta( {  {\vec  \jmath} \over {|\psi{|^2}}} ) 
  \equiv  \int K + \oint \eta(k_g)
\label{energiegeom}
\end{equation}
 The boundary correction is a  function $\eta$
 of the geodesic 
curvature. The results obtained in the preceeding section  
 show that the geodesic curvature is
 given by $n -  {\phi_e}$ for a cylindrically symmetric system and  using 
the appropriate  approximations we determined
 the  function $\eta$ as being an  even  fourth order polynomial in 
 the geodesic curvature.

 This geometric interpretation makes us believe that an expression such as
 (\ref{energiegeom}) is fairly general. It could be well suited,
 as an Ansatz, to describe finite systems which are known to have
 a topological description in the infinite limit (for instance,
 a suitable  generalisation of (\ref{energiegeom}) to other symmetry group like SU(2) 
 could  describe some phases of superfluids in a bounded domain).

\section {comparison with the experimental results}

In order to compare our results to the experimental data
  given in \cite{deo}, 
 we shall consider the limit
 where the radius $R$ is larger than $\lambda$ and $\xi$,
 (typically,  
 $ R \sim 10 \lambda$ is considered experimentally \cite{geim,deo}).
  The thickness of the system is assumed to be small  enough
 so that 
 we can neglect variations of both the magnetic
 field and the order parameter along the cylinder axis.
  We shall prove that within these approximations,  
  the  expression (\ref{gibbs}) captures
   the main  features observed experimentally i.e. the
 behaviour of the 
magnetization at low fields (before the first discontinuity), the
 periodicity and the 
linear behaviour between the successive jumps
 and  provides  a fairly good
 quantitative agreement.

 The magnetization curve in fig.3
 (plotted for the ratio $ {\lambda \over R} = 0.14 $,
 chosen arbitrarily) 
 agrees qualitatively with both the numerical and experimental curves of fig.3 
 in \cite{deo}.
 Besides, taking the  experimental parameters of \cite{deo},
 namely $ R = 1.2 {\mu}m $ and 
 $ \lambda (T) = 84 nm$ 
 at $T = 0.4 K$, we compute from our expressions $ {H_1} = 25 G $ and 
 $ {\Delta}H = 4.6 G .$ These 
 values agree with the results of \cite{deo}
 to within a few percent. We emphasize 
 that $ H_1$ scales like  $1 \over R$,
 whereas $ {\Delta}H$ scales like $ 1 \over {R^2}$ in accordance with 
 \cite{geim}. We calculate the ratio of the magnetization
 jumps to the maximum value of $M$ to be 
 0.20 as compared to a measured value of 0.22. The total
 number of jumps scales like 
 ${R^2}$ and the upper critical field is independent of $R$ in our theory.
 This to agrees fairly 
 well with the experimental data \cite{geim,deo}. We emphasize that
 this quantitative agreement 
 cannot be obtained from the linearized Ginzburg-Landau equations.
 
\section{Conclusion}

We have investigated the problems of the existence and the
 stability of vortices in a two-dimensional 
bounded superconducting system. To that purpose, we
 have used the Ginzburg-Landau energy functional
at the special dual  point characterized by the value
 $\kappa = {1 \over \sqrt 2}$ which 
corresponds, for an infinite system, to a superconductor
 between type I and type II. 
We have shown for that case  that it is still possible
 to obtain vortex solutions at the 
thermodynamic equilibrium. In contrast to the Bogomol'nyi 
solution obtained for the infinite plane, where
 there is no definite value for the number $n$ of vortices, there is
 a selection mechanism for  
 a finite billiard in an external magnetic field 
 that allows to compute the number 
 of vortices as a function of the applied field.
  Our reasoning is  based on the existence of a
 contour  $\Gamma$ which allows to 
 separate  the system in two  parts, the bulk 
containing  the 
vortices and the edge with  screening currents.
 As such  the system might be viewed as 
a kind of two-dimensional Josephson junction or weak superfluid link 
 \cite{anderson}.
 Vortices  enter  (or are expelled from)  the superconductor
 through $\Gamma$, if the applied 
 magnetic field is  increased (or decreased).

Although we considered the special case of the dual point, our
 analysis provides a satisfactory 
quantitative description of the behaviour observed
 experimentally on such small superconducting 
billiards at least for the regime where the density
 of vortices is low enough i.e. for 
small applied magnetic fields. For higher fields, the
 expression (\ref{gibbs}) does not describe 
properly the tail of the magnetization curve where both
 the periodicity $\Delta H$ and the 
amplitude $\Delta M$ of the jumps decrease and
 eventually vanish. In this regime, the vortices 
interact both between themselves and with the
 edge currents. This regime can be studied by a 
perturbative analysis around the dual  point
 \cite{am}.

 So far,  we have studied   only equilibrium states.  On the 
basis of  numerical simulations of the time-dependent Ginzburg-Landau 
equations \cite{argentins}, we  propose a  mechanism
 for vortex nucleation.
At each discontinuity  of the winding number, namely when 
${\phi_e} - {R \over {{2 \sqrt 2}\lambda}}$ is a half integer,
 there is a nodal
 line, joining the center of the system to its boundary, along
 which the order 
parameter $\psi$ vanishes and where  its phase is  ill defined.
  This might be interpreted as an opening 
of the ring of the screening currents which  allows  a flux line of
 the external field to enter the system. In this case,  we expect  the 
 contour  $\Gamma$  to  coincide   with the 
boundary of the system.
The existence of such a  nodal line has been discussed in the context 
of  Ginzburg-Landau equations and for 
 the related  Aharonov-Bohm problem  of a magnetic 
flux line piercing either the infinite plane \cite{berry}
 or a finite domain
 \cite{am,helffer}.

{\bf Acknowledgment}
It is our pleasure to thank A.Ori and T.N.Tomaras for useful and 
enlightening discussions.
This work is supported in part by a grant from the Israel Academy of
Sciences and by the fund for promotion of Research at the Technion.
K.M. acknowledges support  by the Lady Davies foundation.
E.A. ackowledges the very kind hospitality of the Laboratoire de Physique 
des Solides and the LPTMS at the university of Paris (Orsay).

\begin{figure}
{\hspace*{-0.2cm}\psfig{figure=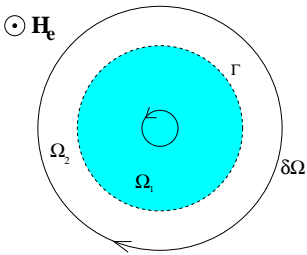,height=4cm,angle=270}}
{\vspace*{.13in}}
\caption{Schematic set-up of the total system with the two subdomains 
$\Omega_1$ and $\Omega_2$ separated by the contour $\Gamma$. }
\label{fig:1}

\end{figure}

\begin{figure}
{\hspace*{-0.2cm}\psfig{figure=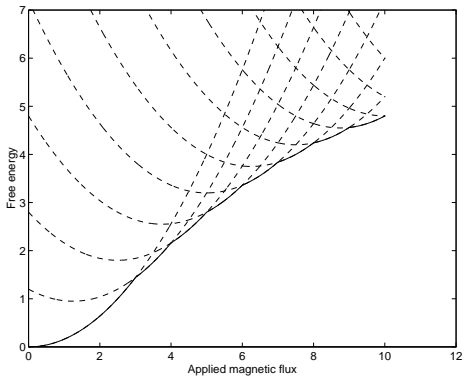,height=5cm,angle=360}}
{\vspace*{.13in}}
\caption{${\cal F}(n, {\phi_e})$ given by (1) plotted as a function of
 the applied magnetic flux $\phi_e$ for various values of the integer $n$
 and for  $\lambda / R = 0.14$.
 The free energy is the 
 envelop of the ensemble of parabolas.}
\label{fig:1}
\end{figure}

\begin{figure}
{\hspace*{-0.2cm}\psfig{figure=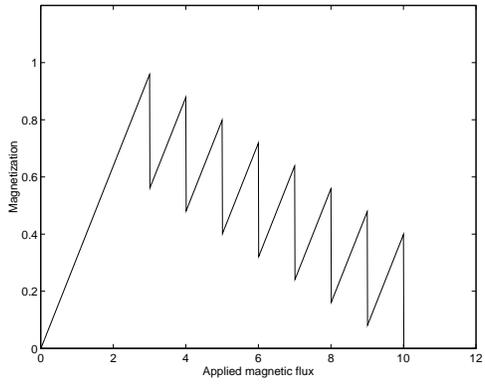,height=5cm,angle=360}}
{\vspace*{.13in}}
\caption{The magnetization $- M$ as given by (15)
 as a function of the applied magnetic flux $\phi_e$ and
 for
 $\lambda/ R = 0.14$  .}
\label{fig:2}
\end{figure}


\begin{thebibliography}{article}
\bibitem{remark} More generally,
 we claim \cite{am} that, even in the absence of  cylindrical symmetry,
 the result (10) remains valid because there   is still
  a contour  $\Gamma$, surrounding the vortices,
 such that the current density  at each point of $\Gamma$ is
 either zero, or orthogonal to $\Gamma$.
\end{thebibliography}
\end{document}